\documentclass[11pt]{cernrep}
\usepackage{graphicx}
\usepackage{cite,./mcite}
\begin{document}
\title{A multi-channel Poissonian model for multi-parton scatterings}
\author{Daniele Treleani}
\institute{Dipartimento di Fisica Teorica
dell'Universit\`a di Trieste, \\INFN and ICTP,
        Trieste, I 34014 Italy }
\maketitle
\begin{abstract}
Multiple parton interactions are typically implemented in Montecarlo codes by assuming a Poissonian distribution of collisions with average number depending on the impact parameter. A possible generalization, which links the process to hadronic diffraction, is shortly discussed. 
\end{abstract}

\section{The simplest Poissonian model}

\noindent A standard way to introduce multiple parton interactions in Montecarlo codes is to assume a Poissonian distribution of multiple parton collisions, with  average number depending on the value of the impact parameter. The motivations were discussed long ago in several articles\cite{Capella:1986cm}\cite{Sjostrand:1987su}\cite{Ametller:1987ru}: One introduces the three dimensional parton density
$D(x,b)$, namely the average number of partons with a given
momentum fraction $x$ and with transverse coordinate $b$ (the
dependence on flavor and on the resolution of the process is
understood) and one makes the simplifying assumption $D(x,b)=G(x)f(b)$, with $G(x)$ the usual
parton distribution function and $f(b)$ normalized to one. The
inclusive cross section for large $p_t$ parton production $\sigma_S$ may hence be expressed as

\begin{eqnarray}
\sigma_S=\int_{p_t^c}G(x)\hat{\sigma}(x,x')G(x') dxdx'=\int_{p_t^c}G(x)f(b) \hat{\sigma}(x,x')G(x')f(b-\beta)
         d^2 bd^2\beta dxdx'
\end{eqnarray}

\noindent where $p_t^c$ is a cutoff introduced to distinguish hard
and soft parton collisions and $\beta$ the hadronic impact parameter. The expression allows a simple geometrical interpretation, given the large momentum exchange which localizes the partonic interaction inside the overlap volume of the two hadrons. 

\noindent Neglecting all correlations in the multi-parton distributions, the inclusive cross
section for a double parton scattering $\sigma_D$ is analogously given by

\begin{eqnarray}
\sigma_D&=&{1\over 2!}\int_{p_t^c}G(x_1)f(b_1)\hat{\sigma}(x_1,x_1')G(x_1')f(b_1-\beta)d^2b_1dx_1dx_1'\times\nonumber\\
&&\qquad\qquad\times
         G(x_2)f(b_2)\hat{\sigma}(x_2,x_2') G(x_2')f(b_2-\beta)
         d^2b_2dx_2dx_2'd^2\beta\nonumber\\
         &=&\int{1\over 2!}\Big(\int_{p_t^c} G(x)f(b)\hat{\sigma}(x,x')G(x')f(b-\beta)d^2bdxdx'\Big)^2d^2\beta
\end{eqnarray}

\noindent
which may be readily generalized to the case of the inclusive cross section for $N$-parton scatterings $\sigma_N$:

\begin{eqnarray}
\sigma_N=\int{1\over N!}\Big(\int_{p_t^c}G(x)f(b)\hat{\sigma}(x,x')G(x')f(b-\beta)d^2bdxdx'\Big)^Nd^2\beta
\end{eqnarray}

\noindent The cross sections are divergent for
$p_t^c\to0$. The unitarity problem is solved by normalizing the
integrand which,
being dimensionless, may be understood as the probability to have a $N$th parton collision process in a inelastic event:

\begin{equation}
\int_{p_t^c}G(x)f(b)\hat{\sigma}(x,x')G(x')f(b-\beta)d^2bdxdx'\equiv \sigma_SF(\beta),\qquad{\bigl(\sigma_SF(\beta)\bigr)^N\over N!}
 e^{-\sigma_SF(\beta)}\equiv P_N(\beta)\end{equation}

\noindent here $P_N(\beta)$ the probability of having $N$ parton collisions in a
hadronic interaction at impact parameter $\beta$. By summing all probabilities one
obtains the hard cross section $\sigma_{hard}$, namely the
contribution to the inelastic cross section due to all events with
{\it at least} one parton collision with momentum transfer greater
than the cutoff $p_t^c$:

\begin{eqnarray}
\sigma_{hard}=\sum_{N=1}^{\infty}\int P_N(\beta)d^2\beta =\sum_{N=1}^{\infty}\int d^2\beta{\bigl(\sigma_SF(\beta)\bigr)^N\over N!}
 e^{-\sigma_SF(\beta)}=\int d^2\beta\Bigl[1-e^{-\sigma_SF(\beta)}\Bigr]
\end{eqnarray}

\noindent Notice that
$\sigma_{hard}$ is finite in the infrared limit, which allows to
express the inelastic cross section as $\sigma_{inel}=\sigma_{soft}+\sigma_{hard}$ with $\sigma_{soft}$ the soft contribution, the two
terms $\sigma_{soft}$ and $\sigma_{hard}$ being defined
through the cutoff in the momentum exchanged at parton level,
$p_t^c$.

\noindent
An important property is that the single parton scattering inclusive
cross section is related to the average number of parton collisions.
One has:

\begin{eqnarray}
\langle N\rangle\sigma_{hard}=\int d^2\beta\sum_{N=1}^{\infty}NP_N(\beta)=\int d^2\beta\sum_{N=1}^{\infty}
{N\bigl[\sigma_SF(\beta)\bigr]^N\over N!}
 e^{-\sigma_SF(\beta)}
=\int d^2\beta \sigma_SF(\beta)=\sigma_S
\end{eqnarray}

\noindent
and more in general one may write:

\begin{eqnarray}
{\langle N(N-1)\dots(N-K+1)\rangle\over K!}\sigma_{hard}
&=&\int d^2\beta \sum_{N=1}^{\infty}
{N(N-1)\dots(N-K+1)\over K!}P_N(\beta)\nonumber\\
&=&\int d^2\beta {1\over K!}\bigl[\sigma_SF(\beta)\bigr]^K=\sigma_K
\end{eqnarray}

\noindent One should stress that the relations between $\sigma_S$ and $\langle N\rangle$
and between $\sigma_K$ and $\langle N(N-1)\dots(N-K+1)\rangle$
do not hold only in the case of the simplest
Poissonian model. It can be shown that the validity is indeed much more general\cite{Calucci:1991qq}\cite{Calucci:1997ii}.

\section{The multi-channel Poissonian model}

An implicit assumption in the Poissonian model is that the hadron density is the same in each interaction. On the other hand the hadron is a dynamical system, which fluctuates in different configurations in a time of the order of the hadron scale, much longer as compared with the time of a hard interaction. Interactions may hence take place while hadrons occupy various configurations, even significantly different as compared with the average hadronic configuration. A measure of the size of the phenomenon is given by hadronic diffraction. 

The multichannel eikonal model\cite{Gotsman:1999ri} allows a simple description of hadronic diffraction. In the multichannel model the hadron state $\psi_h$ is represented as a superposition of  eigenstates $\phi_i$ of the $T$-matrix, while the interaction is described by eikonalized multi-Pomeron exchanges.

\begin{equation}
\psi_h=\sum_i\alpha_i\phi_i
\end{equation}

\noindent The eigenstates of the $T$-matrix can only be absorbed or scatter elastically and the cross sections of the physically observed states $\psi_h$ can be expressed by the combinations of the cross sections between the eigenstates $\phi_i$ and $\phi_j$ as shown below:
 
\begin{eqnarray}
\sigma_{tot}&=&\sum_{i,j}|\alpha_i|^2|\alpha_j|^2\sigma_{tot}^{ij}\cr\nonumber
\sigma_{el}+\sigma_{sd}+\sigma_{dd}&=&\sum_{i,j}|\alpha_i|^2|\alpha_j|^2\sigma_{el}^{ij}\cr\nonumber
\sigma_{in}&=&\sum_{i,j}|\alpha_i|^2|\alpha_j|^2\sigma_{in}^{ij}
\end{eqnarray}

\noindent In a single Pomeron exchange, one may distinguish between hard and soft inelastic interactions, according with the presence or absence of large $p_t$
partons in the final state. One may thus write:

\begin{equation}\sigma^{ij}=\sigma_J^{ij}+\sigma_S^{ij}\end{equation}

\noindent where the labels $J$ or $S$ correspond to the presence or absence of large $p_t$ partons in the final state. One hence obtains the following expression of the hard cross section\cite{Treleani:2007gi}: 

\begin{eqnarray}
\sigma_{hard}&=&\sum_{i,j}|\alpha_i|^2|\alpha_j|^2\sigma_{hard}^{ij}=\sum_{i,j}|\alpha_i|^2|\alpha_j|^2\int d^2\beta\Bigl[1-e^{-\sigma_J^{ij}(\beta)}\Bigr]\cr&=&\sum_{i,j,N}|\alpha_i|^2|\alpha_j|^2\int d^2\beta{\bigl(\sigma_J^{ij}(\beta)\bigr)^N\over N!}
e^{-\sigma_J^{ij}(\beta)}
\end{eqnarray}

\noindent which, being a superposition of Poissonians, represents the natural generalization of the result of the simplest Poissonian model. 

\noindent The easiest implementation of the multi-channel eikonal model is in the case of two eigenstates: 

\begin{equation}
\psi_h={1\over\sqrt 2}\phi_1+{1\over\sqrt 2}\phi_2
\end{equation}

\noindent One obtains:

\begin{equation}
\sigma_{hard}={1\over 4}\sigma_{hard}^{11}+{1\over 2}\sigma_{hard}^{12}+{1\over 4}\sigma_{hard}^{22}
\end{equation}

\noindent while the $N$-parton scattering inclusive cross section $\sigma_{N}$ is given by:

\begin{equation}
\sigma_{N}={\sigma_{S}^{N}\over N!}\Biggl\{{1\over 4}\int \big[F_{11}(\beta)\big]^Nd^2\beta+{1\over 2}\int \big[F_{12}(\beta)\big]^Nd^2\beta+{1\over 4}\int \big[F_{22}(\beta)\big]^Nd^2\beta\Biggr\}
\end{equation}

\noindent where  $F_{ij}$ are the superpositions of the parton densities of the different eigenstates  $\phi_i$ and $\phi_j$. The case of gaussian parton densities is particularly simple. One has

\begin{equation}
F_{ij}(\beta)={1\over \pi(R_i^2+R_j^2)}\times{\rm exp}\Bigl({-\beta^2\over R_i^2+R_j^2}\Bigr)
\end{equation}

\noindent One may take for the radii of the two parton densities  $R_1^2=R^2/2$ and $R_2^2=3R^2/2$, in such a way that $R^2$ is the average hadron size. With this choice the variance of the distribution is $\omega_{\sigma}=1/4$, in agreement with the analysis of  \cite{Blaettel:1993ah}.
The explicit expression of the inclusive cross section $\sigma_{N}$ is: 

\begin{equation}
\sigma_{N}={\sigma_{S}^{N}\over NN!(\pi R^2)^{N-1}}\Biggl\{{1\over 4}\Biggl({1\over {1\over 2}+{1\over 2}}\Biggr)^{N-1}+{1\over 2}\Biggl({1\over {1\over 2}+{3\over 2}}\Biggr)^{N-1}+{1\over 4}\Biggl({1\over {3\over 2}+{3\over 2}}\Biggr)^{N-1}\Biggr\}
\end{equation}

\begin{figure}[h]
\vspace{0cm}
\includegraphics[width=18cm]{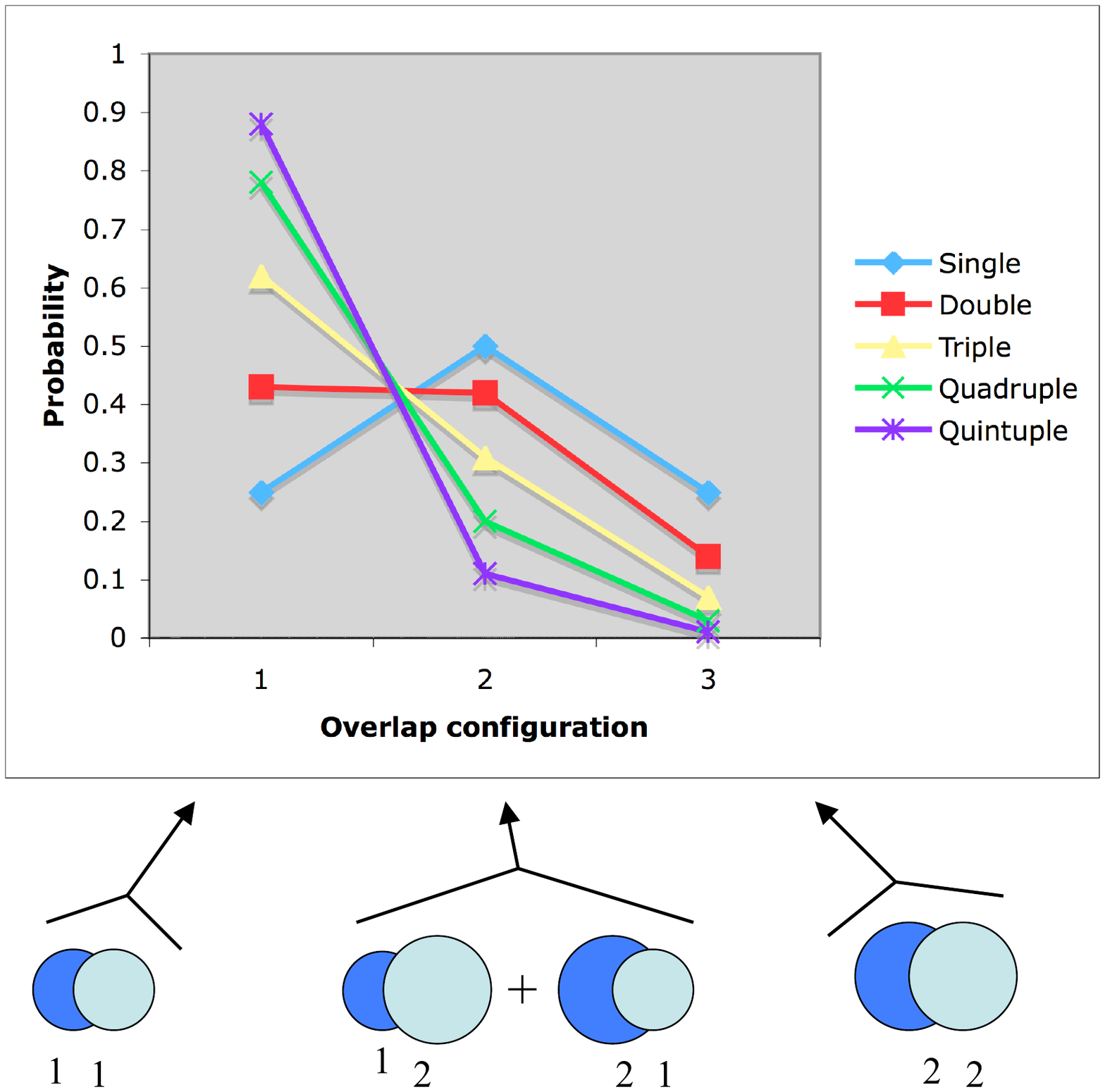}
\vspace{0cm}
\caption{Relative contributions to the inclusive cross section $\sigma_N$ of the overlaps between the different hadronic configurations for $1\le N\le 5$ }
\label{fig:overlaps}
\end{figure}

\noindent In the figure the relative weights of the overlaps between the various configurations are shown for different inclusive cross sections $\sigma_N$. In the case of a single collision all four different configurations contribute with the same weight. When $N$ grows the contribution of the overlap between the most compact and dense configurations becomes increasingly important and, for $N$=5, it acconts for almost $90\%$ of the cross section. 

Notice that the result obtained in the multi-channel eikonal model {\it is very different} with respect to the result obtained when terms with various transverse sizes are introduced directlly in the hadronic parton density of the simplest Poissonian model. In  PYTHIA\cite{Sjostrand:1987su}\cite{Sjostrand:2006za} the hadron density is represented by the sum of two gaussians with same weight and different size. The overlap function is hence given by

\begin{equation}
{1\over 4}F_{11}(\beta)+{1\over 2} F_{12}(\beta)+{1\over 4}F_{22}(\beta)
\end{equation}
 
\noindent where $F_{ij}$ are given by Eq.14, with $R_1$ and $R_2$ the radii of the two gaussians used to construct the actual hadronic parton density. The resulting expression of the inclusive cross sections is

\begin{equation}
\sigma_{N}={\sigma_{S}^{N}\over N!}\int \Big[{1\over 4}F_{11}(\beta)+{1\over 2} F_{12}(\beta)+{1\over 4}F_{22}(\beta)\Big]^Nd^2\beta
\end{equation}

\noindent which should be compared with the inclusive cross section derived in the two-states eikonal model (expression in Eq.13). 

\section{Concluding remarks}

In the present note it has been shown how the importance of small size hadronic configurations is emphasized  by geometry in the multi-parton inclusive cross sections $\sigma_N$ at large $N$. Here one has assumed that the transverse fluctuations of the hadron do not affect its parton content. In the two-states-model of hadronic diffraction one needs however to enhance the strength of the Pomeron coupling between diffractive eigenstates with small radii, in order to fit the available data on elastic, inelastic, single and double diffractive cross sections\cite{Gotsman:2007ac}, which corresponds to an increase of the parton content when the hadron occupies a configuration with small transverse size. In the analysis\cite{Blaettel:1993ah} hadronic diffraction is on the contrary fitted in a model where the number of partons decreases when the hadron occupies small size configurations. While in the former case the enhanced role of small transverse size configurations in multiparton collisions is amplified\cite{Treleani:2007gi}, in the latter it is on the contrary reduced\cite{Frankfurt:2008vi}.

The study of hadronic diffraction and of multiparton scatterings at the LHC may hence provide non trivial informations on the correlation between  the parton content of the hadron and its transverse size. In addition to the measurements of hadronic diffraction and of multi-jets cross sections in hadron-hadron collisions, an important handle, to gain a better insight into this aspect of the hadron structure, may be represented by the measurement of multi-jets cross sections in hadron-nucleus collisions, where a model independent separation of the longitudinal and transverse parton correlations is, in principle, possible\cite{Strikman:2001gz}. 

%------------------------------------------------------------------------------
%       Bibliography
%------------------------------------------------------------------------------
\bibliographystyle{heralhc} 
{\raggedright
\bibliography{heralhc2008}
}

\end{document}